\documentclass[11pt]{article}
\usepackage{amsmath,amssymb,amsthm,amsxtra,bbm,bbold,bm,epsfig,multirow,longtable}
\usepackage{xcolor,ulem,subfigure,hyperref,cite}

\def\thefootnote{\fnsymbol{footnote}} 
\allowdisplaybreaks

\definecolor{darkpink}{RGB}{219, 48, 122}

\pdfoutput=1

\begin{document}

\vspace{0.2cm}

\begin{center} 
{\Large\bf Symmetries and stabilisers in modular invariant flavour models}
\end{center}

\begin{center}
{\bf Ivo de Medeiros Varzielas$^1$}\footnote{Email: \tt ivo.de@udo.edu}
,
{\bf Miguel Levy$^1$}\footnote{Email: \tt miguelplevy@ist.utl.pt}
and 
{\bf Ye-Ling Zhou$^2$}\footnote{Email: \tt ye-ling.zhou@soton.ac.uk}
\\\vspace{5mm}
{$^1$CFTP, Departamento de F\'{\i}sica, Instituto Superior T\'{e}cnico,}\\
Universidade de Lisboa,
Avenida Rovisco Pais 1, 1049 Lisboa, Portugal \\
{$^2$ School of Physics and Astronomy, University of Southampton,\\
Southampton SO17 1BJ, United Kingdom } 
\\
\end{center}

\vspace{1.5cm} 

\begin{abstract} 
The idea of modular invariance provides a novel explanation of flavour mixing.
Within the context of finite modular symmetries $\Gamma_N$ and for a given element $\gamma \in \Gamma_N$, we present an algorithm for finding stabilisers (specific values for moduli fields $\tau_\gamma$ which remain unchanged under the action associated to $\gamma$). We then employ this algorithm to find all stabilisers for each element of finite modular groups for $N=2$ to $5$, namely, $\Gamma_2\simeq S_3$, $\Gamma_3\simeq A_4$, $\Gamma_4\simeq S_4$ and $\Gamma_5\simeq A_5$. These stabilisers then leave preserved a specific cyclic subgroup of $\Gamma_N$. This is of interest to build models of fermionic mixing where each fermionic sector preserves a separate residual symmetry.
\end{abstract}
\begin{flushleft}
\hspace{0.8cm} PACS number(s): 14.60.Pq, 11.30.Hv, 12.60.Fr \\
\hspace{0.8cm} Keywords: Lepton flavour mixing, flavour symmetry
\end{flushleft}

\def\thefootnote{\arabic{footnote}}
\setcounter{footnote}{0}

\newpage

\section{Introduction}

Non-Abelian discrete symmetries were introduced to understand the theoretical origin of large lepton mixing angles observed in neutrino oscillation experiments. A popular approach is that the lepton flavour mixing is realised by the spontaneous symmetry breaking (SSB) of discrete flavour symmetries \cite{Altarelli:2005yp,Altarelli:2005yx}. 
This approach requires the introduction of new scalars called flavons. They get vacuum expectation values (VEVs), leading to SSB of the symmetry, and Yukawa couplings appear as the effective consequence of the VEVs of flavons (see, e.g., \cite{King:2017guk,Xing:2019vks,Feruglio:2019ktm} for some recent reviews).

A different approach based on modular invariance \cite{Ferrara:1989bc,Ferrara:1989qb} was recently proposed in \cite{Feruglio:2017spp}. In the new approach, a finite modular symmetry $\Gamma_N$ (for level $N$) and a modulus field $\tau$ are assumed in models of leptonic masses and mixing, and the Yukawa couplings appear as modular forms of $\tau$ with an even modular weight  \cite{Criado:2018thu}. This idea was generalised to multiple modular symmetries with moduli fields in \cite{deMedeirosVarzielas:2019cyj}. 
The modular invariance approach enables models with few or no flavons. 
Interesting models which are trying to explain neutrino masses and lepton mixing have been constructed in different finite modular symmetries, 
specifically in $\Gamma_2\simeq S_3$~\cite{Kobayashi:2018vbk, Kobayashi:2018wkl}, 
$\Gamma_3\simeq A_4$~\cite{Feruglio:2017spp, Criado:2018thu, Kobayashi:2018scp, Okada:2018yrn, Kobayashi:2018wkl, Novichkov:2018yse, Ding:2019zxk, Zhang:2019ngf, Wang:2019xbo}, 
$\Gamma_4\simeq S_4$~\cite{Penedo:2018nmg, Novichkov:2018ovf, deMedeirosVarzielas:2019cyj, King:2019vhv, Wang:2019ovr,Wang:2020dbp}, and
$\Gamma_5\simeq A_5$~\cite{Novichkov:2018nkm, Ding:2019xna}. 
Discussion was extended to the modular symmetry with a higher level $N=7$, $\Gamma_7 \simeq PSL_2(Z_7) \simeq \Sigma(168)$ which includes complex triplet representations \cite{Ding:2020msi}. 
The modular invariance approach was also generalised to include odd-weight modular forms  \cite{Liu:2019khw,Liu:2020akv,Novichkov:2020eep} and half-integer modular forms \cite{Liu:2020msy}, which can be arranged into irreducible representations of the homogeneous
finite modular groups $\Gamma'_N$ and finite metaplectic group $\widetilde{\Gamma}_{4N}$, respectively. 
Modular symmetries have also been applied to the quark flavours \cite{Kobayashi:2018wkl,Okada:2018yrn, Okada:2019uoy,King:2020qaj}.
An important ingredient of the modular invariance approach is the modulus field $\tau$ or moduli fields as considered in the context of multiple modular symmetries \cite{deMedeirosVarzielas:2019cyj}. Alternative to role of flavons in former flavour model construction, the modulus field play the essential rule in the SSB of flavour symmetry. 

Although it is still not clear how the modulus gains a VEV in flavour models, some particularly interesting values of $\tau$, which are invariant under particular modular transformations, have taken particularly relevance in the literature. They are called stabilisers of the relevant modular transformation. These values may play important role in modular symmetry breaking and special mixing pattern. 
Such an idea was briefly discussed in \cite{Novichkov:2018yse} and \cite{Novichkov:2018ovf}, based on $\Gamma_3 \simeq A_4$ and $\Gamma_4 \simeq S_4$, respectively. Special values for $\tau$ might be obtained e.g. through orbifolding \cite{deAnda:2018ecu}. 
In \cite{deMedeirosVarzielas:2019cyj}, it has been explicitly proven that modular forms at a stabiliser
preserve a residual subgroup of the finite modular symmetry and are eigenvectors of representation matrices of the relevant elements in the subgroup. 
Realistic models were constructed in the framework of multiple $S_4$ modular symmetries \cite{deMedeirosVarzielas:2019cyj, King:2019vhv}, where stabilisers are crucial to achieve TM$_1$ mixing. The new TM$_1$ mixing patters are more predictive than those predicted in the flavon approach due to the intrinsic property of modular forms \cite{King:2019vhv}. 

In this paper we introduce an algorithm to find stabilisers and then perform a systematic scan to find stabilisers for each element of finite modular groups for $N=2$ to $5$, i.e. $\Gamma_{2,3,4,5}$.
A recent work \cite{Gui-JunDing:2019wap} has similarly studied stabilisers (referred to as fixed points) in $\Gamma_{3,4}$ ($A_4, S_4$). Our results agree with those presented therein. A comparison of their methodology with ours will appear in Section \ref{sec:scan}. One of our main new results is that we present the complete list of stabilizers for $\Gamma_{2,5}$. The work of \cite{Gui-JunDing:2019wap} is also useful to showcase the relevance of these results for model-building.

In Section \ref{sec:ms_stab} we briefly review the framework of modular symmetries and stabilisers in general terms.
Section \ref{sec:scan} starts with an explanation of the algorithm, then its systematical application. We present the results in figures showing the stabilisers in the domains of the respective modular symmetries and we list them in tables displaying each group element and respective stabilisers.
In Section \ref{sec:conc} we conclude.

\section{Modular symmetry and stabilisers \label{sec:ms_stab}}

\subsection{Modular symmetry}

The modular group $\overline{\Gamma}$ is made of elements acting on the complex modulus $\tau$ (${\rm Im}(\tau)>0$) as linear fractional transformations:
\begin{eqnarray} \label{eq:modular_transformation}
\gamma: \tau \to \gamma \tau = \frac{a \tau + b}{c \tau + d}\,,
\end{eqnarray}
where $a, b, c, d$ are integers and the condition $ad-bc=1$ is satisfied. 
Each element of $\overline{\Gamma}$ can be represented by a two by two matrix\footnote{This matrix need not be a unitary matrix.} up to an overall sign difference. Then, $\overline{\Gamma}$ is expressed to be 
\begin{eqnarray}
\overline{\Gamma} = \left\{ \begin{pmatrix} a & b \\ c & d \end{pmatrix} / (\pm \mathbf{1})\,,\;\; a, b, c, d \in \mathbb{Z}, \;\;  ad-bc=1  \right\} \,.
\end{eqnarray}
The modular group $\overline{\Gamma}$ is isomorphic to the projective spacial linear group $PSL(2,\mathbb{Z}) = SL(2,\mathbb{Z})/\mathbb{Z}_2$. 
It has two generators, $S_\tau$ and $T_\tau$, satisfying $S_\tau^2 = (S_\tau T_\tau)^3 = e$. For later convenience and brevity, we define the order 3 generator $C_\tau \equiv S_\tau T_\tau$. The generators act on the modulus $\tau$ as
\begin{eqnarray}
S_\tau: \tau \to -\frac{1}{\tau} \,, \hspace{1cm}
T_\tau: \tau \to \tau + 1\,,
\end{eqnarray}
respectively. Representing them by two by two matrices, we obtain 
\begin{eqnarray}
S_\tau=\begin{pmatrix} 0 & 1 \\ -1 & 0 \end{pmatrix}\,, \hspace{1cm}
T_\tau=\begin{pmatrix} 1 & 1 \\ 0 & 1 \end{pmatrix} \,.
\end{eqnarray}

$\overline{\Gamma}$ is a discrete and infinite group. By requiring $a, d = 1~({\rm mod}~N)$ and $b, c =  0~({\rm mod}~N)$ with $N=2, 3, 4, 5, \cdots$, we obtain a subset of $\overline{\Gamma}$ which is also an infinite group and labelled as $\overline{\Gamma}(N)$, 
\begin{eqnarray}
\overline{\Gamma}(N) = \left\{ \begin{pmatrix} n_aN+1 & n_bN \\ n_cN & n_dN+1 \end{pmatrix}  / (\pm \mathbf{1})\,,~ ~ n_a, n_b, n_c, n_d \in \mathbb{Z},
 ~~  (n_a n_d- n_b n_c) N + n_a + n_d = 0  \right\} \,.
\label{eq:modN}
\end{eqnarray}
The quotient group $\overline{\Gamma}/\overline{\Gamma}(N)$ is labelled as $\Gamma_N$. It is finite and called finite modular group. 
The finite modular group $\Gamma_N$ for $N \leq 5$ can also be obtained by imposing conditions $S_\tau^2 = (S_\tau T_\tau)^3 = T_\tau^N = e$,\footnote{For $N>5$, additional conditions have to imposed to make the group finite. e.g., $(S_\tau T_\tau^3)^4 = e$ for $N=7$ \cite{Ding:2020msi}.} where the last condition can be achieved by identifying $\tau=\tau+N$ in the upper complex plane.
Note that once $\tau=\tau+N$ is imposed, $\tau' = \frac{-1}{\tau} =  \frac{-1}{\tau+N} = \frac{-\tau'}{N\tau' -1}$ is automatically satisfied.
In this way, the definition of $\Gamma_N$ can be written as
\begin{eqnarray}\label{eq:modN2}
\Gamma_N = \left\{ \begin{pmatrix} a & b \\ c & d \end{pmatrix}  / (\pm \mathbf{1})\,,~ a, b, c, d \in \mathbb{Z}_N, ~~  ad-bc=1 \right\} \,.
\end{eqnarray}

For $N$ taking some small number, $\Gamma_N$ is isomorphic to a permutation group, 
in particular, $\Gamma_2 \simeq S_3$, $\Gamma_3 \simeq A_4$, $\Gamma_4 \simeq S_4$ and $\Gamma_5 \simeq A_5$ \cite{deAdelhartToorop:2011re}. In this work we will tend to refer to the groups in $\Gamma_N$ notation. The fundamental domains for each of these will appear in Section \ref{sec:scan}.

\subsection{Stabilisers and residual modular symmetries \label{sec:residual}}

We now discuss the target space of modular symmetry, as we are interested in finding the full stabilisers of a finite modular group. 

We label the fundamental domain of $\overline{\Gamma}$ and $\overline{\Gamma}(N)$ as $\mathcal{D}$ and $\mathcal{D}(N)$, respectively. The fundamental domain $\mathcal{D}$ is defined as follows. Given a point $\tau$ in the upper complex plane, acting all modular transformations of $\overline{\Gamma}$ on $\tau$ forms an orbit of the point $\tau$. The fundamental domain $\mathcal{D}$ of $\bar{\Gamma}$ represents a minimal connected region of $\tau$. In this region, every orbit of $\overline{\Gamma}$ intersects $\mathcal{D}$ in at least one point, and each orbit that intersects the interior of $\overline{\Gamma}$ intersects $\mathcal{D}$ in no more than one point. Similarly, one defines the fundamental domain $\mathcal{D}(N)$ of $\overline{\Gamma}(N)$.

 Acting $\overline{\Gamma}$ on $\mathcal{D}$ generates $\mathcal{C} \equiv \mathbb{C}_+ \cup \{ \rm cusps \}$, namely, the upper complex plane (${\rm Im}(\tau)>0$) with cusps on the real axis. On the other hand, acting $\overline{\Gamma}(N)$ on $\mathcal{D}(N)$ generates the same space. Therefore, we have 
\begin{eqnarray}
\mathcal{C} = \overline{\Gamma} \mathcal{D} = \overline{\Gamma}(N) \mathcal{D}(N) \,.
\end{eqnarray}
Since $\Gamma_N$ represents the quotient group $\overline{\Gamma}/\overline{\Gamma}(N)$, we further have $\overline{\Gamma} \mathcal{D} = \overline{\Gamma}(N) \Gamma_N \mathcal{D}$. Comparing with the former equation, we obtain
\begin{eqnarray}
\label{eq:Domain}
\mathcal{D}(N) = \Gamma_N \mathcal{D} \,.
\end{eqnarray}
$\mathcal{D}(N)$ forms the full target space of $\Gamma_N$. Any transformation $\gamma \in \Gamma_N$ acting on $\mathcal{D}(N)$ leaves $\mathcal{D}(N)$ invariant, $\gamma\mathcal{D}(N) = \mathcal{D}(N)$. On the other hand, acting each element $\gamma$ of $\Gamma_N$ on the fundemental domain of $\overline{\Gamma}$ generates another fundamental domain, i.e., $\gamma \mathcal{D} \neq \mathcal{D}$.

Given an element $\gamma$ in the modular group $\Gamma_N$, 
a stabiliser of $\gamma$, which may not be unique, corresponds to a fixed point $\tau_\gamma$ either in the interior or on the boundary of the fundamental domain $\mathcal{D}(N)$ \footnote{We do not need to consider the full upper complex plane. } which satisfies $\gamma \tau_\gamma = \tau_\gamma$. Some of the properties satisfied by stabilisers are discussed below. 
\begin{itemize}
\item 
Since each orbit of $\overline{\Gamma}$ intersects the interior of $\mathcal{D}$ in no more than one point, a stabiliser of $\gamma \in \Gamma_N$ should be located only on either an edge or cusp of one fundamental domain of $\bar{\Gamma}$. 

\item
A stabiliser of $\gamma$ is also a stabiliser of $\gamma^2, \gamma^3, \cdots$, since $\gamma^2 \tau_\gamma = \gamma \tau_\gamma = \tau_\gamma$. 
Therefore, once the modular field $\tau$ gain a VEV at such a stabiliser, $\langle \tau \rangle = \tau_\gamma$, an Abelian residual modular symmetry $Z_\gamma = \{\mathbf{1}, \gamma, \gamma^2, \cdots \}$ is preserved. 

\item Given a stabiliser  $\tau_\gamma$ of $\gamma$, $\gamma_1 \tau_\gamma$ is a stabiliser of the conjugate $\gamma_1 \gamma \gamma_1^{-1}$. This is simply proven as $\gamma_1  \gamma \gamma_1^{-1} \gamma_1 \tau_\gamma = \gamma_1  \gamma \tau_\gamma = \gamma_1 \tau_\gamma$. A specific consequence is that if there is an element $\gamma_1$ which is not equal to $\gamma$ but permutes with $\gamma$, $\gamma = \gamma_1 \gamma \gamma_1^{-1}$, and then both $\tau_\gamma$ and $\gamma_1 \tau_\gamma$ are stabilisers of $\gamma$. 
Therefore, one modular transformation of $\Gamma_N$ may have several different stabilisers in $\mathcal{D}(N)$. 

\end{itemize}

Given $a$, $b$, $c$ and $d$ for any element $\gamma \in \Gamma_N$ with $a$, $b$, $c$ and $d$ being integers and $ad-bc=1$, the most general $2 \times 2$ matrix of $\gamma$ should be written as
\begin{eqnarray}
\gamma = \eta \begin{pmatrix} N n_a +a & N n_b +b \\ N n_c +c & N n_d +d \end{pmatrix} \,,
\end{eqnarray} 
where $n_a$, $n_b$, $n_c$ and $n_d$ are any integers and satisfy $N n_a n_d + a n_d + d n_a = N n_b n_c + b n_c + c n_b$ and $\eta= \pm 1$. 
Stabilisers of $\gamma$ can be obtained by solving the following equations
\begin{eqnarray}
\frac{(N n_a +a)\tau + N n_b +b}{(N n_c +c) \tau + N n_d +d}=\tau \,.
\end{eqnarray}
Solutions of $\tau$ must be located in $\mathcal{D}(N)$, which is always achieved by selecting a typical set of integers  $n_a$, $n_b$, $n_c$ and $n_d$. Using these conditions, we are able to obtain full lists of stabilisers for all modular transformations of $\Gamma_N$. 

Here we show stabilisers for the generator $S_\tau$ of $\Gamma_2$, where $S_\tau^2=e$. The element $S_\tau$ can be represented as 
\begin{eqnarray} 
S_\tau =\begin{pmatrix} 0 & 1 \\ -1 & 0 \end{pmatrix} = \begin{pmatrix} 0 & 1 \\ -1 & 2 \end{pmatrix} = \begin{pmatrix} 0 & 1 \\ -1 & -2 \end{pmatrix} \,, 
\label{eq:op_2d}
\end{eqnarray}
where we have taken $n_a=n_b=n_c=0$, and $n_d=0, 1, -1$, from left to right.
For these possibilities, we solve $S_\tau \tau = \tau$ and obtain 
\begin{eqnarray} 
\tau_{S_{\tau},1}= i\,,\quad 
\tau_{S_{\tau},2}= 1\,, \quad
\tau_{S_{\tau},3}= -1\,,
\label{eq:multiple2D}
\end{eqnarray}
where $\tau_{S_{\tau},1}$, $\tau_{S_{\tau},2}$ are different stabilisers of $S_\tau$ in $\mathcal{D}(2)$. Given the relation $\tau=\tau+N$, we find that $\tau_{S_{\tau},2} = \tau_{S_{\tau},3}$. It should be obvious that some choices of $n_a, n_b, n_c, n_d$ lead to stabilisers outside $\mathcal{D}(N)$, making it such that not all sets of integers are fruitful, and making this list finite.

Modular forms of a given weight $k$ and for a given level $N$ are simply holomorphic functions (of $\tau$) which transform in a specific way under $\Gamma_N$:
\begin{eqnarray} \label{eq:modform}
Y_I(\gamma \tau) = (c\tau + d)^{-2k} Y_I(\tau) \,.
\end{eqnarray} 
Modular forms are particularly important, as they are the building blocks of models based on invariance under a modular symmetry, similar to irreducible representations. Fields and couplings are assigned as modular forms and invariant terms are built from combining them in an appropriate way.

It is obvious that acting $\gamma$ on a modular form at its stabiliser leaves the modular form invariant, i.e., 
\begin{eqnarray}
\gamma: Y_I(\tau_\gamma) \to Y_I(\gamma \tau_\gamma) = Y_I(\tau_\gamma)\,. 
\end{eqnarray}
Following the standard transformation property Eq.(\ref{eq:modform}), $Y_I(\gamma \tau_\gamma) = (c \tau_\gamma + d )^{2 k} \rho_I (\gamma) Y_I (\tau_\gamma)$, 
we obtain 
\begin{eqnarray} \label{eq:yukawa_eigenvector}
\rho_I(\gamma) Y_I(\tau_\gamma) = (c\tau_\gamma + d)^{-2k} Y_I(\tau_\gamma) \,, 
\end{eqnarray} 
where $\rho_I(\gamma)$ is the representation matrix of $\gamma$. 
This equation lead us to the following important properties for the stabiliser and the modular form \cite{deMedeirosVarzielas:2019cyj}:
\begin{itemize}

\item A modular form multiplet at a stabiliser, that is $Y_I(\tau_\gamma)$, is an eigenvector of the representation matrix $\rho_I(\gamma)$ with corresponding eigenvalue $(c\tau_\gamma + d)^{-2k}$.

\item The stabiliser $\tau_\gamma$ satisfies $|c\tau_\gamma + d| = 1$ since $(c\tau_\gamma + d)^{-2k}$ is an eigenvalue of a unitary matrix. 
\end{itemize}
A special case is that when $(c\tau_\gamma + d)^{-2k}=1$ is satisfied, $\rho(\gamma) Y(\tau_\gamma) = Y(\tau_\gamma)$, and we recover the residual flavour symmetry generated by $\gamma$. 
In general, the eigenvalue does not need to be fixed at $1$ in the framework of modular symmetry.

\section{Stabilisers for finite modular groups \label{sec:scan}}

A straightforward way to understand how to find an extensive list of stabilisers, is to make use of disjoint sections of the domain of $\Gamma_N$. In the following, we take $\mathcal{D}$ to be the domain defined as $\{ \tau \in \mathbb{C}: |\tau|>1, \left|\Re(\tau)\right|<1/2 \}$, combined with a suitable choice of boundaries. These disjoint (barring boundaries) regions are obtained by acting all  elements of $\Gamma_N$ on $\mathcal{D}$, and span the fundamental domain of the group (cf. Eq.~\eqref{eq:Domain}). As such, all points in $\gamma \mathcal{D}$ for $\gamma \in \Gamma_N$ are bijectively related to points in $\mathcal{D}$, in a one-to-one mapping. This is the property we exploit to find an extensive list of all stabilisers in $\Gamma_N$.

Since, excluding boundaries, acting any element $\gamma$ on $\tau$ will transform it from $\mathcal{D_\tau}$ to $\gamma\mathcal{D_{\tau}}$, then for any non-boundary point to be a stabiliser ($\gamma \tau =\tau$), it would require $\gamma \mathcal{D_\tau}=\mathcal{D_\tau}$, and thus $\gamma=e$. On the other hand, for a point $\tau$ on the boundary of $\mathcal{D}$, it is possible to act $\gamma$ such that the point remains on $\mathcal{D}$, if $\mathcal{D}$ and $\gamma \mathcal{D}$ share a border. This is succintly put in the first properties satisfied by stabilisers, shown in Section~\ref{sec:residual}. Hence, if it is possible to find $\gamma_1 \tau = \gamma_2 \tau$, with $\gamma_1 \neq \gamma_2$, for any boundary point, then $\gamma_1 \gamma_2^{-1} \tau=\tau$. 

Exploring the boundaries of $\mathcal{D}$, there are four well-known stabilisers \cite{Novichkov:2018ovf} (where they appear with different notation):
\begin{itemize}
\item $\tau_1=i$, $\gamma_1=S_\tau$, $\gamma_2=e$
\item $\tau_2=\frac{1}{2}+\frac{i \sqrt{3}}{2}$ , $\gamma_1= T_\tau$, $\gamma_2 = S_\tau$
\item $\tau_3=-\frac{1}{2}+\frac{i \sqrt{3}}{2}$, $\gamma_1= T^{N-1}_\tau$, $\gamma_2 = S_\tau$
\item $\tau_4=i \infty$, $\gamma_1=T_\tau$, $\gamma_2=e$
\end{itemize}

Given the stabilisers in $\mathcal{D}$, it is possible to propagate these onto the remaining sections, by acting all elements of $\Gamma_N$ on the stabilisers. Since these will span the entire (fundamental) domain of $\Gamma_N$, a list of stabilisers arises, containing all non-equivalent possibilities. It is noteworthy to say that, while the specific methodology holds for any choice of Domain, the upper complex plane does has a many-to-one mapping to the domain of $\Gamma_N$. Namely, due to the relation $T_\tau^N = e$, we have
\begin{equation}
\tau=\frac{\tau}{1 + n_1 N\tau} + n_2 N, \qquad n_1, n_2 \in \mathbb{Z} .
\label{eq:degeneracy}
\end{equation}
Thus, there are an infinite number of points in the upper complex plane which are equivalent to each other, for $\Gamma_N$. As such, the lists obtained show the stabilisers $\tau$, where $\tau$ belongs to our chosen domain. That is not to say that there are no other points in $\mathcal{C}$ which stabilise a certain element of $\Gamma_N$, but rather that those elements are equivalent to one of the stabilisers given here. One may also naively think that a certain point is not stabilised by its corresponding element. Again, this is due to the redundancy of points in $\mathcal{C}$, shown in Eq. \eqref{eq:degeneracy}.

\begin{figure}
\centering
\includegraphics[width=.4\linewidth]{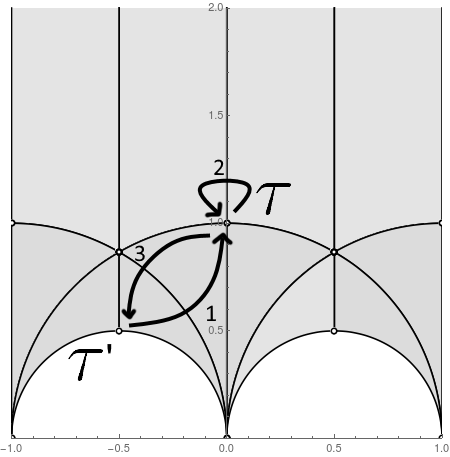}
\caption{\small An example of the applied methodology to find the stabilisers of $\Gamma_N$. The example shown is for $\Gamma_2$, where the arrows denote the actions of different elements, $\gamma^{-1}$, $\gamma_i$, $\gamma$, for 1,2,3 respectively, following the convention of the text.}
\label{fig-method}
\end{figure}

Lastly, after finding the full list of stabilisers, one needs to find the element $\gamma$ for which $\gamma\tau=\tau$. The methodology for the whole process is a straightforward 3-step computation:
\begin{enumerate}
\item Take $\tau=\tau_i$, where $\tau_i = \gamma_i \tau_i, i=1, ..., 4$ is a stabiliser of $\mathcal{D}$;
\item Act $\gamma$ on $\tau$: $\tau'=\gamma \tau$. Compute $\gamma^{-1}$;
\item The element that stabilises $\tau'$ is given by $\gamma^{-1} \gamma_i\, \gamma$.
\end{enumerate}
The idea behind this simple process is exemplified in Figure~\ref{fig-method}. By comparing with the methodology exposed above, we see that we act $\gamma$ on $\tau=i$, to find $\tau'$. Then, the element that stabilises $\tau'$ is $\gamma^{-1} \gamma_i \, \gamma$, where each action is represented by an arrow. Namely, $\gamma^{-1}$, $\gamma_i$, and $\gamma$ are shown by arrows 1, 2, and 3, respectively. 

This procedure is also presented in \cite{Gui-JunDing:2019wap}. Here, we complement this procedure with the second property of stabilisers shown in Section \ref{sec:ms_stab} (if $\tau$ is a stabiliser of $\gamma$, then it is also a stabiliser of powers of $\gamma$) to achieve a more complete list of stabilisers for each element.

In the following subsections, we show the fundamental domain $\mathcal{D}(N)$\footnote{Domains at the top (e.g. $\mathcal{D}$) continue to complex $+ \infty$, whereas domains represented at the edges overlap and points should not be double counted. For simplicity, we represent both boundaries in the figures.} of $\bar{\Gamma}(N)$ for $N=2,3,4,5$. The complete lists of stabilisers for all elements of $\Gamma_N$ are found both in the domain, and separately in a table with its corresponding stabilising element. A given element can be stabilised by more than one stabiliser, and further, a stabiliser for a specific element necessarily also stabilises powers of that element and therefore preserves the associated cyclic subgroup. The elements are listed according to their conjugacy classes. 
In general, stabilisers are located at special points in the complex plane, namely in intersections or midway points in the domains of each group. This makes intuitive sense particularly when considering the association of these $N \leq 5$ groups to geometric objects (triangle, tetrahedron, cube or octahedron, dodecahedron or icosahedron) and the possible mapping between the fundamental domains in complex space and these objects.
Given that there are redundancies in the boundaries of $\mathcal{D}(N)$, we choose to keep only one choice for the stabiliser in the table (we opt for the right-most $\tau$, i.e. the one with largest real part), and also show a list of equivalencies between relevant boundary points for $\Gamma_N$.

Even though these finite modular groups can be generated by a minimal set of 2 elements $S_\tau$ and $T_\tau$, it was convenient for us to identify each group element through 3 (related) generators $S_\tau$, $T_\tau$ and $C_\tau = S_\tau T_\tau$ (note that while the order of $T_\tau$ is $N$, for $\Gamma_N$, the order of $S_\tau$ is 2 and the order of $C_\tau$ is 3, regardless). For a given irreducible representation (the doublet of $\Gamma_2$ and triplets of the remaining groups) we present in a specific basis the elements $S_\tau$ and $T_\tau$ in the respective subsection, as well as an example of what is the modular form for that representation at a given stabiliser.

\subsection{$\Gamma_2$ and its stabilisers}

In the framework of modular symmetry, $\Gamma_2$ is obtained by fixing $N=2$, such that we have $S_\tau^2  = (S_\tau T_\tau)^3 = T_\tau^2 = e$. $\Gamma_2$ is isomorphic to $S_3$, the group of permutations of 3 objects and the symmetry of the equilateral triangle. We relate the generators $S_\tau$ and $T_\tau$ to a conventional set of generators in cycle notation, e.g. $S_\tau = (12)$ and $T_\tau = (31)$, where the equalities between generators of the modular group and of the cycle notation generators of $S_3$ (or the symmetries of the triangle) are taken in the sense of the isomorphism relating them. The 6 elements are then $\{ e, S_\tau, T_\tau, S_\tau T_\tau, T_\tau S_\tau, T_\tau S_\tau T_\tau \}$ with $S_\tau T_\tau S_\tau = T_\tau S_\tau T_\tau = (23)$. The conjugacy classes are the $\{e\}$, the 3-cycles (3-fold rotations of the triangle) $\{ S_\tau T_\tau, T_\tau S_\tau \} = \{ (123), (321) \}$ and the 2-cycles  (reflections of the triangle) $\{ S_\tau, T_\tau, T_\tau S_\tau T_\tau \} = \{ (12), (31), (23) \}$. 
We recall also our definition of $C_\tau \equiv S_\tau T_\tau$.

\begin{figure}[ht]
\centering
\includegraphics[height=.2\textheight,keepaspectratio]{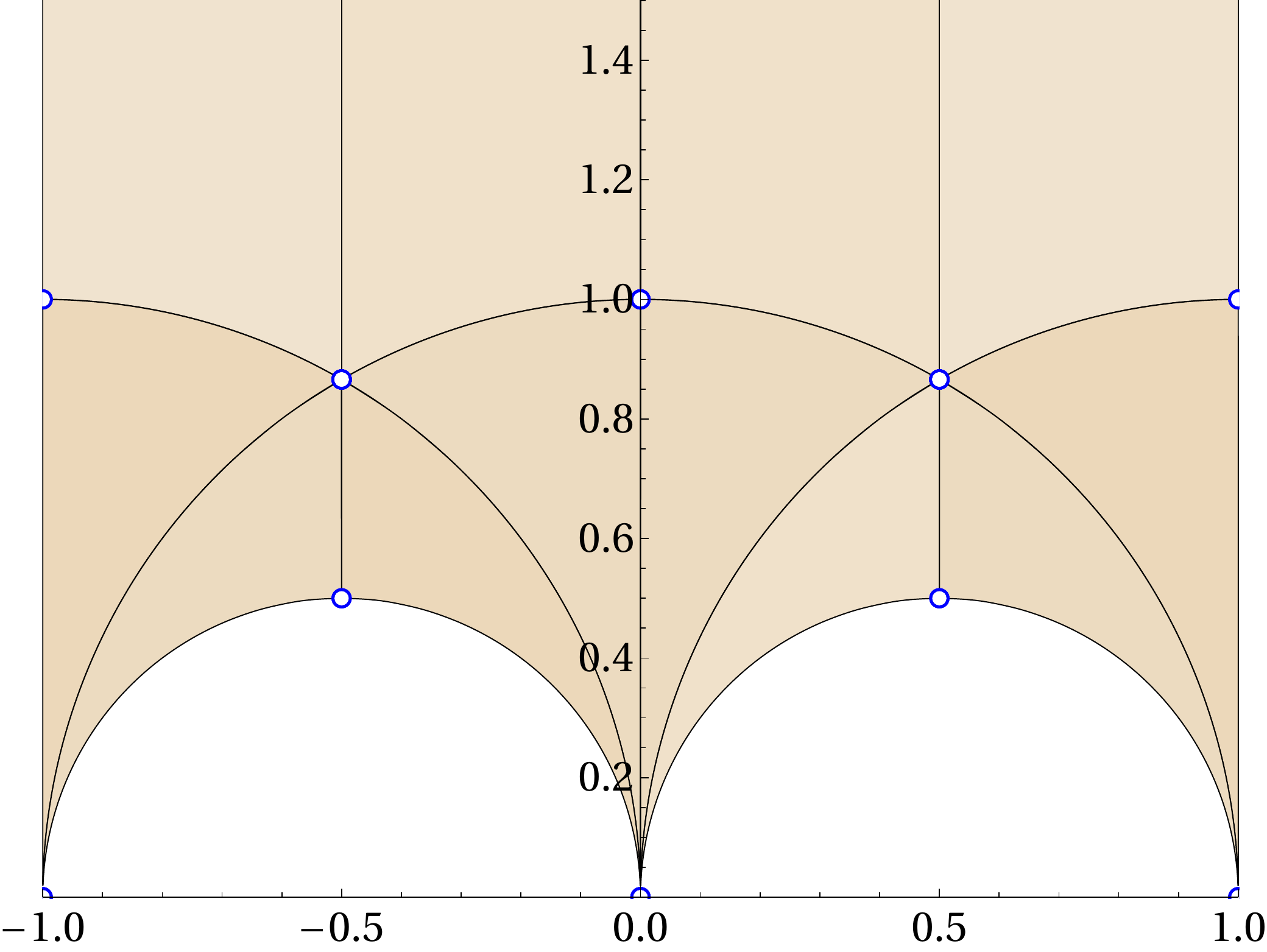}
\caption{ \small \label{fig:G2} The fundamental domain $\mathcal{D}(2)$ of $\bar{\Gamma}(2)$ (i.e., the full target space of $\Gamma_2\simeq S_3$) with the stabilisers of modular transformations of $\Gamma_2$ denoted as dots.}
\end{figure}

We depict the fundamental domain of $\Gamma_2$ and the location of the stabilisers in the complex plane in Figure \ref{fig:G2}.
Table \ref{ta:G2} has a complete list of stabilisers. For $S_3$, the relevant redundancies are:
\begin{eqnarray}
\label{eq:G2-redunds}
\frac{1}{2}+\frac{i}{2} = -\frac{1}{2}+\frac{i}{2} \,, \qquad 1 = -1.
\end{eqnarray}
For the sake of clarity, we show here a proof of the first redundancy shown, since it also helps understand why $\tau = \frac{1}{2}+\frac{i}{2}$ is a stabiliser of $T_\tau$. Although we are adressing this issue specifically for $\Gamma_2$, the reasoning holds for the remaining modular symmetries here shown. 
Let us start with the element $\gamma=S_\tau T_\tau S_\tau T_\tau S_\tau$. It is easily seen that $\gamma$ stabilises $\tau=\frac{1}{2}+\frac{i}{2}$:
\begin{equation}
\frac{1}{2}+\frac{i}{2} \xrightarrow{S_\tau} -1 +i \xrightarrow{T_\tau} i  \xrightarrow{S_\tau} i \xrightarrow{T_\tau} 1+i \xrightarrow{\tau =\tau+N } -1+i \xrightarrow{S_\tau} \frac{1}{2}+\frac{i}{2} \,.
\end{equation}
Additionally, it can be shown that $\gamma= T_\tau$, for the case of $\Gamma_2$:
\begin{equation}
\gamma = (S_\tau T_\tau S_\tau T_\tau S_\tau) (T_\tau T_\tau^{-1}) =(S_\tau T_\tau S_\tau T_\tau S_\tau T_\tau) T_\tau = T_\tau, 
\end{equation}
where we used $T_\tau^{-1}=T_\tau$, and $(S_\tau T_\tau)^3=e$. Hence, we see that, for $\Gamma_2$, $\gamma \tau_\gamma =  -\frac{1}{2}+\frac{i}{2}=\frac{1}{2}+\frac{i}{2}$, and $T_\tau \tau_\gamma = \tau_\gamma$, for $\tau_\gamma=\frac{1}{2}+\frac{i}{2}$.

 This could also be shown using Eq.~\eqref{eq:degeneracy}, by taking $n_1=-1$, $n_2=0$, and obviously $N=2$.
Although this may not always be possible by a single application of Eq.~\eqref{eq:degeneracy}, multiple consecutive applications of this relation would link any two redundant points.

In this way, the table shows $\tau=(1+i)/2$ (with larger real part than $\tau=(-1+i)/2$), and this list of equivalencies complements the table, by stating these points are identical in $\Gamma_2$, and thus any of the two are effectively stabilisers of the corresponding element. As stated above, this game could be endlessly played, since there are an infinite number of redundancies in $\mathcal{C}$. However, here, we restrict ourselves to the redundancies belonging to the fundamental domains (up to the redundancies of the boundaries and cusps, as we discussed earlier) of the respective groups, shown in Figs.~\ref{fig:G2}~to~\ref{fig:G5}.

\LTcapwidth=\linewidth
\begin{table}[h!]
\begin{center}
\begin{tabular}{c|c|c} \hline
	& $\gamma$ & $\tau_\gamma$ \\ \hline \hline
&&\\[-2ex] 
\multirow{3}{*}{$\mathcal{C}_2$} &  $T_\tau C_\tau$ & $0, 1+i$ \\
							& $T_\tau$ & $i \infty, \frac{1}{2}+\frac{i}{2}$ \\
							 & $S_\tau$ & $i,1$
\\ \hline  \hline 
\rule{0pt}{4ex}    \multirow{2}{*}{$\mathcal{C}_3$} &  $T_\tau S_\tau$ & $-\frac{1}{2}+\frac{i \sqrt{3}}{2},\frac{1}{2}+\frac{i \sqrt{3}}{2}$\\
							& $C_\tau$ & $-\frac{1}{2}+\frac{i \sqrt{3}}{2},\frac{1}{2}+\frac{i \sqrt{3}}{2}$ \\[-2ex] & &\\\hline
\end{tabular}
\caption{\label{ta:G2} The non-identity elements of $\Gamma_2$ and respective stabilisers.}
\end{center}
\end{table}

For the doublet irreducible representation in a $T_\tau$-diagonal basis, the $S_\tau$ and $T_\tau$ generators take the form
\begin{eqnarray}
\rho_{\mathbf{2}}(S_\tau) = \frac{1}{2} \begin{bmatrix} -1 & \sqrt{3}  \\ \sqrt{3} & 1 \end{bmatrix} \,,\quad
\rho_{\mathbf{2}}(T_\tau) = \begin{bmatrix} 1 & 0 \\ 0 & -1 \end{bmatrix} \,.
\end{eqnarray} 
where here and in following subsection we use square brackets for representation matrices to distinguish from the $2 \times 2$ operators acting in the upper complex plane such as \eqref{eq:op_2d}.
For the doublet of $\Gamma_2$ ($S_3$), modular forms at stabilisers for $S_\tau$ and $T_\tau$ take the form:
\begin{eqnarray}
Y_{\mathbf{2}}(\tau_{S_\tau}) \propto 
\begin{bmatrix} \frac{-\sqrt{3}}{2} \\ \frac{1}{2} \end{bmatrix} \,,~
\begin{bmatrix} \frac{1}{2} \\ \frac{\sqrt{3}}{2} \end{bmatrix} \,, \quad
Y_{\mathbf{2}}(\tau_{T_\tau}) \propto 
\begin{bmatrix} 1 \\ 0 \end{bmatrix} \,,~
\begin{bmatrix} 0 \\ 1 \end{bmatrix} \,.
\end{eqnarray}
These forms are directly determined following the discussion after Eq.~\eqref{eq:yukawa_eigenvector}. Only the overall factor cannot be determined.

\subsection{$\Gamma_3$ and its stabilisers}

\begin{figure}[ht]
\centering
\includegraphics[height=.2\textheight,keepaspectratio]{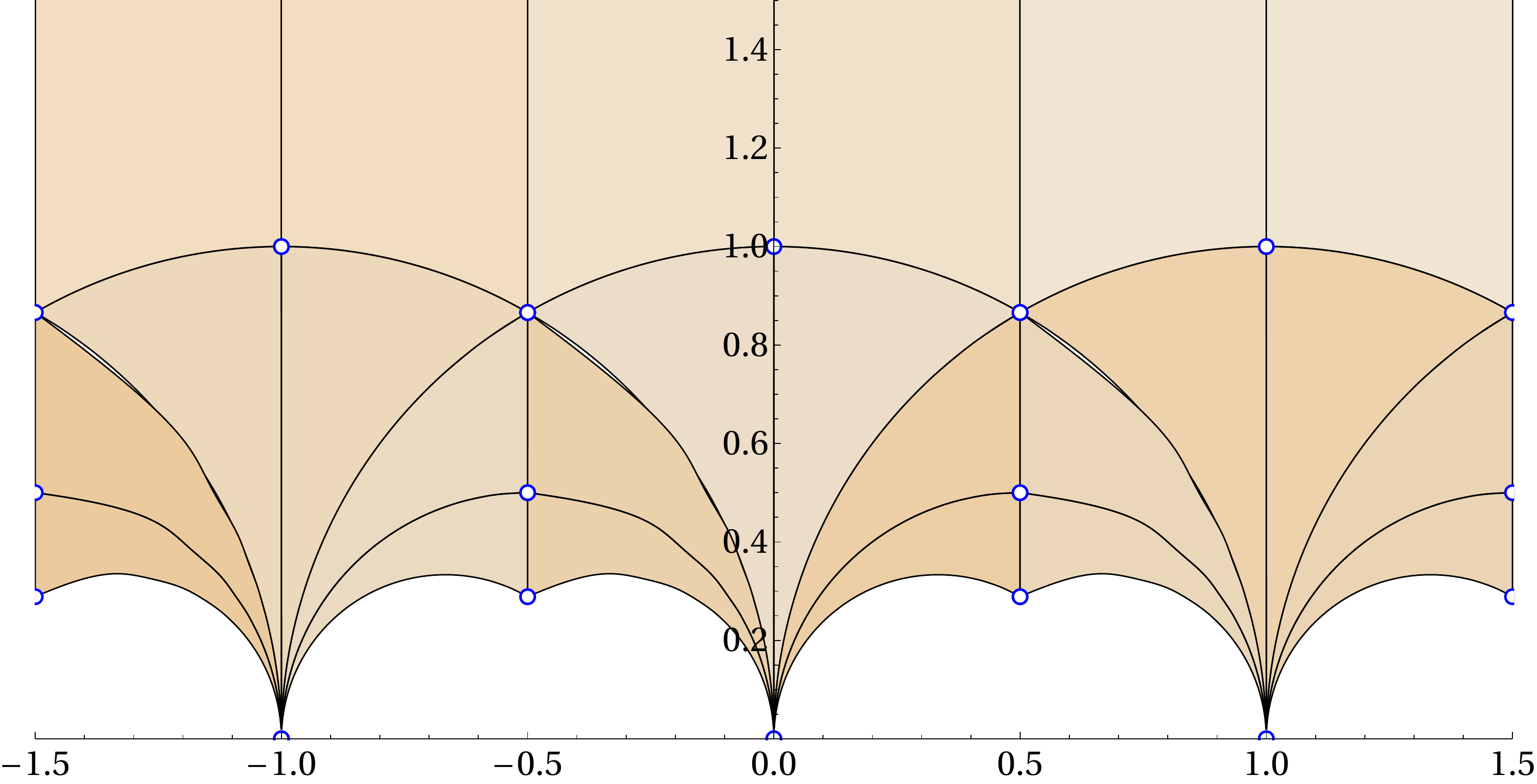}
\caption{\label{fig:G3} The fundamental domain $\mathcal{D}(3)$ of $\bar{\Gamma}(3)$ with the stabilisers of modular transformations of $\Gamma_3$ denoted as dots.}
\end{figure}

$\Gamma_3$ has presentation $S_\tau^2  = (S_\tau T_\tau)^3 = T_\tau^3=e$. It is isomorphic to $A_4$, the group of even permutations of four objects and the symmetry group of the tetrahedron.
For $\Gamma_{3}$, $S_\tau$ can be interpreted geometrically as a reflection and $T_\tau$ as a 3-fold rotation.
We consider 3 generators $S_\tau$, $T_\tau$ and $C_\tau = S_\tau T_\tau$ as described before. The list of equivalencies between the relevant boundary points of the domain shown in Fig.~\ref{fig:G3} is:
\begin{eqnarray}
&\frac{3}{2}+\frac{i}{2 \sqrt{3}} = \frac{1}{2}+\frac{i}{2 \sqrt{3}} = -\frac{1}{2}+\frac{i}{2 \sqrt{3}} = -\frac{3}{2}+\frac{i}{2 \sqrt{3}} \,, \nonumber\\
&\frac{3}{2}+\frac{i}{2} =-\frac{3}{2}+\frac{i}{2} \,, \qquad \frac{3}{2}+\frac{i \sqrt{3}}{2} =-\frac{3}{2}+\frac{i\sqrt{3}}{2} \,.
\end{eqnarray}
These relations complement Table~\ref{ta:G3}.

\begin{table}[h!]
\begin{center}
\begin{tabular}{c|c|c}
\hline
	& $\gamma$ & $\tau_\gamma$ \\ \hline \hline
&&\\[-2ex] 
\multirow{4}{*}{$\mathcal{C}_2$} &
   $C_\tau^2$ & $-\frac{1}{2}+\frac{i \sqrt{3}}{2}, 1$ \\
& $T_\tau^2$ & $i \infty,\frac{3}{2}+\frac{i}{2 \sqrt{3}}$ \\
& $T_\tau C_\tau$ & $0, \frac{3}{2}+\frac{i \sqrt{3}}{2}$ \\
& $C_\tau T_\tau$ & $-1,\frac{1}{2}+\frac{i \sqrt{3}}{2}$ 
\\[-2ex] & &
\\ \hline  \hline 
&&\\[-2ex] 
\multirow{4}{*}{$\mathcal{C}_3$} &
   $C_\tau$ & $-\frac{1}{2}+\frac{i \sqrt{3}}{2},1$ \\
& $T_\tau$ & $i \infty, \frac{3}{2}+\frac{i}{2 \sqrt{3}}$ \\
& $C_\tau S_\tau$ & $0,\frac{3}{2}+\frac{i \sqrt{3}}{2}$ \\
& $T_\tau S_\tau$ & $-1,\frac{1}{2}+\frac{i \sqrt{3}}{2}$ \\[-2ex] & &
\\ \hline  \hline 
&&\\[-2ex] 
\multirow{3}{*}{$\mathcal{C}_4$} & 
   $T_\tau^2 C_\tau$ & $-1+i,\frac{1}{2}+\frac{i}{2}$ \\
& $S_\tau$ & $i,\frac{3}{2}+\frac{i}{2}$ \\
& $T_\tau C_\tau T_\tau$ & $-\frac{1}{2}+\frac{i}{2}, 1+i$ \\[-2ex] 
&&\\
\hline
\end{tabular}
\caption{\label{ta:G3} The non-identity elements of $\Gamma_3$ and respective stabilisers.}
\end{center}
\end{table}

The generators $S_\tau$ and $T_\tau$ for the triplet of $A_4$ in a $T_\tau$-diagonal basis have representation matrices:
\begin{eqnarray}
\rho_{\mathbf{3}}(S_\tau) = \frac{1}{3} \begin{bmatrix} -1 & 2 & 2 \\ 2 & -1 & 2 \\ 2 & 2 & -1 \end{bmatrix} \,, \quad
\rho_{\mathbf{3}}(T_\tau) =\begin{bmatrix} 1 & 0 & 0 \\ 0 & \omega^2 & 0 \\ 0 & 0 & \omega \end{bmatrix} \,.
\end{eqnarray} 
Note that this particular choice for the generators has been taken in the literature (see e.g. \cite{Novichkov:2018yse}).
Following the discussion after Eq.~\eqref{eq:yukawa_eigenvector}, we obtain modular forms at stabilisers for $S_\tau$ and $T_\tau$ as
\begin{eqnarray} \label{eq:A4_modular}
&&Y_{\mathbf{3}}(\tau_{S_\tau}) \propto 
\begin{bmatrix}1 \\ 1 \\ 1 \end{bmatrix} \,,~
x \begin{bmatrix} 2 \\ -1 \\ -1 \end{bmatrix} + 
y \begin{bmatrix} 0 \\ 1 \\ -1 \end{bmatrix}\,, \quad
Y_{\mathbf{3}}(\tau_{T_\tau}) \propto 
\begin{bmatrix} 1 \\ 0 \\ 0 \end{bmatrix} \,,~
\begin{bmatrix} 0 \\ 1 \\ 0 \end{bmatrix} \,,~
\begin{bmatrix} 0 \\ 0 \\ 1 \end{bmatrix} \,.
\end{eqnarray}
Here, since $\rho_{\mathbf{3}}(S_\tau)$ has degenerate eigenvalues, $[2,-1,-1]^T$ and $[0,1,-1]^T$ and any of their linear combinations are eigenvectors of $\rho_{\mathbf{3}}(S_\tau)$. To further determine the coefficients $x$ and $y$,  we have to consider either correlations of modular forms (e.g., $Y_2^2 + 2 Y_1 Y_3 = 0$ for weight $k=2$ \cite{Feruglio:2017spp}) or explicit expressions of modular forms.

\subsection{$\Gamma_4$ and its stabilisers}

$\Gamma_4$ is isomorphic to $S_4$, which is the group of all permutations of four objects, and the symmetry group of the cube and of the octahedron. Here, $S_\tau$ can be interpreted geometrically as a reflection whereas $T_\tau$ can be interpreted as a 4-fold rotation.
In the framework of modular symmetry, the $\Gamma_4$ modular group is obtained in the series of $\Gamma_N$ by fixing $N=4$. In other words, its generators satisfy $S_\tau^2  = (S_\tau T_\tau)^3 = T_\tau^4 =e$. 
In former works, it is common to use three generators $S$, $T$ and $U$, which satisfy $S^2 = T^3 = U^2 = (ST)^3 = (SU)^2 = (TU)^2 = e$, to generate $S_4$. These generators can be represented by $S_\tau$ and $T_\tau$ as
\begin{eqnarray}
S = T_\tau^2 \,,~
T = S_\tau T_\tau \,,~
U = T_\tau S_\tau T_\tau^2 S_\tau \,. 
\label{eq:S4gens}
\end{eqnarray}
In the upper complex plane with the requirement $\tau = \tau +4$, $S$, $T$ and $U$ can be represented by two by two matrices such as
\begin{eqnarray} \label{eq:STU}
S=\begin{pmatrix} 1 & 2 \\ 0 & 1 \end{pmatrix}\,, ~
T=\begin{pmatrix} 0 & 1 \\ -1 & -1 \end{pmatrix} \,, ~
U=\begin{pmatrix} 1 & -1 \\ 2 & -1 \end{pmatrix} \,.
\end{eqnarray}
Due to the identification in Eq.~\eqref{eq:modN2}, these representation matrices are not unique. 
It is convenient to write out another three elements of $S_4$, $TS = S_\tau T_\tau^{-1} $, $ST = T_\tau S_\tau T_\tau^{-1} S_\tau$ and $STS = T_\tau^{-1} S_\tau T_\tau S_\tau$. They are order-three elements of $S_4$ and will be used for our later discussion. The two by two representation matrices for them are given by
\begin{eqnarray} \label{eq:TS}
TS=\begin{pmatrix} 0 & 1 \\ -1 & 1 \end{pmatrix}\,,~~
ST=\begin{pmatrix} 2 & -1 \\ 3 & -1 \end{pmatrix}\,,~~
STS=\begin{pmatrix} -2 & -1 \\ 3 & 1 \end{pmatrix}\,.
\end{eqnarray}

\begin{figure}[ht]
\centering
\includegraphics[height=.2\textheight,keepaspectratio]{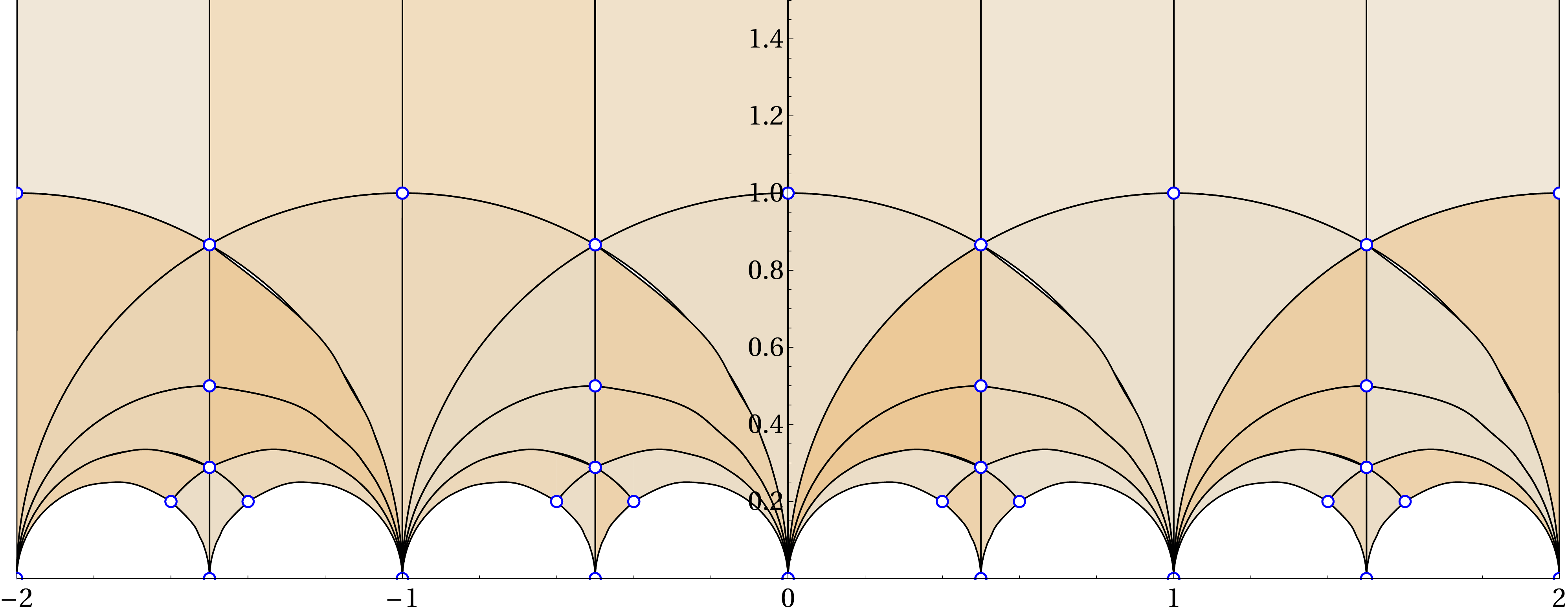}
\caption{\label{fig:G4} The fundamental domain $\mathcal{D}(4)$ of $\bar{\Gamma}(4)$ with the stabilisers of modular transformations of $\Gamma_4$ denoted as dots.}
\end{figure}

We list the target space of $\Gamma_4$, namely, the fundamental domain $\mathcal{D}(4)$, in Fig.~\ref{fig:G4}. 
The list of stabilisers is shown in Table~\ref{ta:G4}, and the redundancies of  the domain shown in Fig.~\ref{fig:G4} are 
\begin{eqnarray}
&2 = -2 \,, \qquad
2+i = -2+i \,, \qquad
\frac{2}{5}+\frac{i}{5} = -\frac{2}{5}+\frac{i}{5} \,, \nonumber\\
&\pm \frac{7}{5}+\frac{i}{5} = \pm \frac{3}{5}+\frac{i}{5} \,, \quad
\frac{8}{5}+\frac{i}{5} = -\frac{8}{5}+\frac{i}{5} \,, \quad
\frac{3}{2} = \frac{1}{2} = -\frac{1}{2} = -\frac{3}{2} \,.
\end{eqnarray}
We note that the $\pm$ in the equation above mean only that the two stabilisers with positive real part are equivalent, and that the two stabilisers with negative real part are equivalent, without further equivalences. These redundancies, both in the chosen target space of $\Gamma_4$ and, as such, for the stabilisers, can be explicitly seen by comparing Fig.~\ref{fig:G4} with the domain and stabilisers shown in \cite{Gui-JunDing:2019wap}.

By inverting the relations of Eq.~\eqref{eq:S4gens}, it is possible to find $S_\tau$ and $T_\tau$ as a function of $S$, $T$, and $U$:
\begin{eqnarray}
S_\tau = S T S U \,,~
T_\tau = S T S U T.
\label{eq:S4genInv}
\end{eqnarray}

For completeness, we show here for the triplet irreducible representations, in a $T$-diagonal basis, the representations matrices for both choices of generators:
\begin{eqnarray}
\rho_{\mathbf{3}^{(\prime)}}(S) = \frac{1}{3} \begin{bmatrix} -1 & 2 & 2 \\ 2 & -1 & 2 \\ 2 & 2 & -1 \end{bmatrix} \,,\quad
\rho_{\mathbf{3}^{(\prime)}}(T)=\begin{bmatrix} 1 & 0 & 0 \\ 0 & \omega^2 & 0 \\ 0 & 0 & \omega \end{bmatrix} \,,  \quad
\rho_{\mathbf{3}^{(\prime)}}(U)=(-)\begin{bmatrix} 1 & 0 & 0 \\ 0 & 0 & 1 \\ 0 & 1 & 0 \end{bmatrix} \,,
\end{eqnarray}
and
\begin{eqnarray}
\rho_{\mathbf{3}^{(\prime)}}(S_\tau) = (-) \frac{1}{3} \begin{bmatrix} -1 & 2 \omega^2 & 2 \omega \\ 2 \omega & 2 & -\omega^2 \\ 2\omega^2 & -\omega & 2  \end{bmatrix} \,, \quad
\rho_{\mathbf{3}^{(\prime)}}(T_\tau) = (-) \frac{1}{3} \begin{bmatrix} -1 & 2 \omega & 2 \omega^2 \\ 2 \omega & 2 \omega^2 & -1 \\ 2 \omega^2 & -1 & 2 \omega  \end{bmatrix} \, ,~ 
\end{eqnarray} 
where $\omega=e^{2 i \pi /3}$.

For the triplet $\mathbf{3}$ of $\Gamma_4$ ($S_4$), modular forms at stabilisers for $S_\tau$ and $T_\tau$ take the form:
\begin{eqnarray}
&&Y_{\mathbf{3}}(\tau_{S_\tau}) \propto
\begin{bmatrix} 2 \\ -\omega \\ -\omega^2 \end{bmatrix} \,,~
x \begin{bmatrix} -\omega \\ 2 \\ 0 \end{bmatrix} + 
y \begin{bmatrix} \omega \\ 0 \\ 2 \end{bmatrix}\,, \quad
Y_{\mathbf{3}}(\tau_{T_\tau}) \propto
\begin{bmatrix} 1 \\ 1 \\ 1 \end{bmatrix} \,,~
\begin{bmatrix} 1-\sqrt{3} \\ \sqrt{3}-2 \\ 1 \end{bmatrix} \,,~
\begin{bmatrix} 1+\sqrt{3} \\ -\sqrt{3}-2 \\ 1 \end{bmatrix} \,,
\end{eqnarray}
where, as in $\Gamma_3$, we use $x,y$ as placeholder normalization factors that can be found (for a specified weight).
We note that $Y_{\mathbf{3}}(\tau_{S})$ and $Y_{\mathbf{3}}(\tau_{T})$ are the same as $Y_{\mathbf{3}}(\tau_{S_\tau})$ and $Y_{\mathbf{3}}(\tau_{T_\tau})$, respectively, in Eq.~\eqref{eq:A4_modular}. In turn, $Y_{\mathbf{3}}(\tau_{U})$ is given by (using $x, y$ factors):
\begin{eqnarray}
&&Y_{\mathbf{3}}(\tau_{U}) \propto
\begin{bmatrix} 0 \\ 1 \\ -1 \end{bmatrix} \,,~
x \begin{bmatrix} 1 \\ 0 \\ 0 \end{bmatrix} + 
y \begin{bmatrix} 0 \\ 1 \\ 1 \end{bmatrix}\,.
\end{eqnarray}

{\centering
\begin{longtable}{c|c|c} \hline
	& $\gamma$ & $\tau_\gamma$ \\ \hline \hline
&&\\[-2ex] 
\multirow{6}{*}{$\mathcal{C}_2$}
& $T_\tau^2 C_\tau T_\tau$ & $\frac{2}{5}+\frac{i}{5},2+i$ \\
& $T_\tau^2 C_\tau S_\tau$ & $1+i, -\frac{3}{5}+\frac{i}{5},$ \\
& $T_\tau C_\tau T_\tau S_\tau$ & $-\frac{3}{2}+\frac{i}{2}, \frac{1}{2}+\frac{i}{2}$ \\
& $S_\tau$ & $i,\frac{8}{5}+\frac{i}{5}$ \\
& $C_\tau T_\tau C_\tau$ & $-\frac{1}{2}+\frac{i}{2},\frac{3}{2}+\frac{i}{2}$ \\
& $C_\tau^2T_\tau$ & $-1+i,\frac{7}{5}+\frac{i}{5}$ \\[-2ex]
&&\\ \hline  \hline 
&&\\[-2ex]
& $C_\tau^2$ & $-\frac{1}{2}+\frac{i \sqrt{3}}{2},\frac{3}{2}+\frac{i}{2 \sqrt{3}}$ \\
& $C_\tau$ & $-\frac{1}{2}+\frac{i \sqrt{3}}{2},\frac{3}{2}+\frac{i}{2 \sqrt{3}}$ \\
& $T_\tau^2 C_\tau$ & $-\frac{3}{2}+\frac{i \sqrt{3}}{2},\frac{1}{2}+\frac{i}{2 \sqrt{3}}$ \\
\multirow{2}{*}{$\mathcal{C}_3$} & $C_\tau T_\tau C_\tau S_\tau$ & $-\frac{3}{2}+\frac{i \sqrt{3}}{2}, \frac{1}{2}+\frac{i}{2 \sqrt{3}}$ \\
& $T_\tau C_\tau T_\tau$ & $-\frac{1}{2}+\frac{i}{2 \sqrt{3}},\frac{3}{2}+\frac{i \sqrt{3}}{2}$ \\
& $C_\tau^2 T_\tau S_\tau$ & $-\frac{1}{2}+\frac{i}{2 \sqrt{3}},\frac{3}{2}+\frac{i \sqrt{3}}{2}$ \\
& $T_\tau C_\tau S_\tau$ & $-\frac{3}{2}+\frac{i}{2 \sqrt{3}}, \frac{1}{2}+\frac{i \sqrt{3}}{2}$ \\
& $T_\tau S_\tau$ & $-\frac{3}{2}+\frac{i}{2 \sqrt{3}},\frac{1}{2}+\frac{i \sqrt{3}}{2}$ \\[-2ex]
&&\\ \hline  \hline 
&&\\[-2ex]  
& $T_\tau^2$ & $i \infty, \frac{3}{2}$ \\
$\mathcal{C}_4$& $C_\tau T_\tau S_\tau$ & $0,2$ \\
& $C_\tau T_\tau C_\tau T_\tau$ & $-1,1$ \\ [-2ex]
&&\\ \hline  \hline 
&&\\[-2ex]
\multirow{6}{*}{$\mathcal{C}_5$} 
& $T_\tau$ & $i \infty, \frac{3}{2}$ \\
& $T_\tau^3$ & $i \infty, \frac{3}{2}$ \\
& $C_\tau S_\tau$ & $0,2$ \\
& $T_\tau C_\tau$ & $0,2$ \\
& $T_\tau^2S_\tau$ & $-1,1$ \\
& $C_\tau T_\tau$ & $-1,1$ \\[-2ex] 
&&\\\hline
\caption{\label{ta:G4} The non-identity elements of $\Gamma_4$ and respective stabilisers.}
\end{longtable}
}

\subsection{$\Gamma_5$ and its stabilisers}

\begin{figure}[h]
\centering
\includegraphics[height=.2\textheight,keepaspectratio]{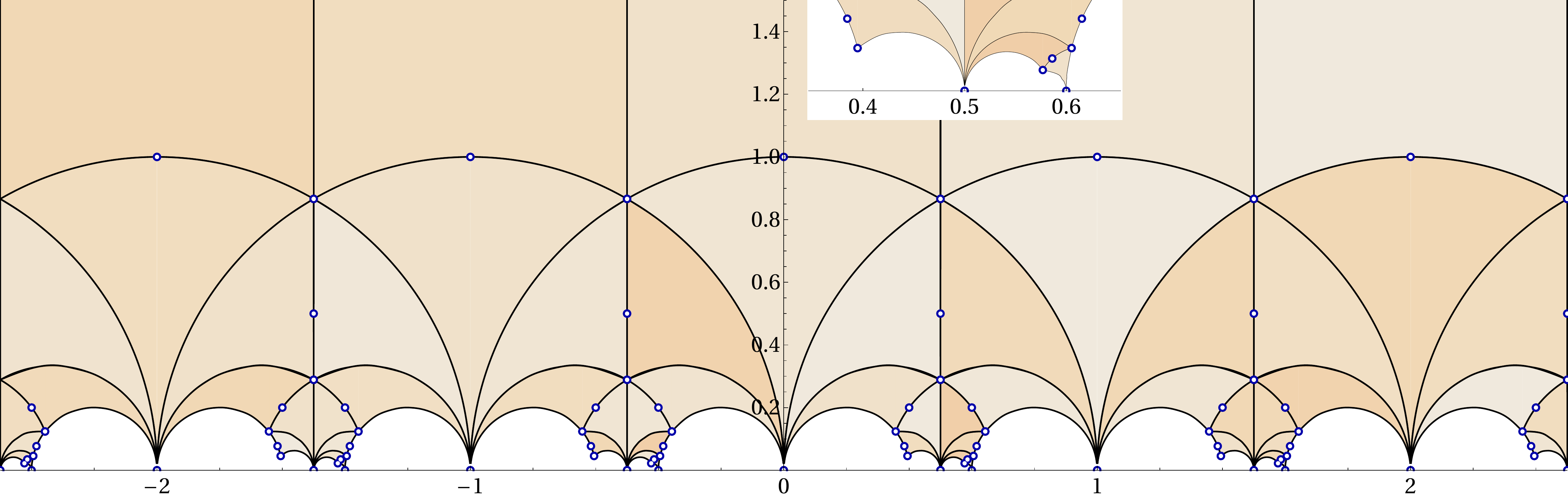}
\caption{\label{fig:G5} The fundamental domain $\mathcal{D}(5)$ of $\bar{\Gamma}(5)$ with the stabilisers of modular transformations of $\Gamma_5$ denoted as dots. In the box on top, we show a zoomed section of the domain around the cusp $\tau=1/2$, where there are many small-sized intricacies. We show only one zoomed section as the remaining areas surrounding the half-integer cusps are identical.}
\end{figure}

$\Gamma_5$ is isomorphic to $A_5$, which is the group of even permutations of five objects and the symmetry group of the dodecahedron and of the icosahedron. The generators satisfy $S_\tau^2  = (S_\tau T_\tau)^3 = T_\tau^5 =e$. $S_\tau$ can be interpreted geometrically as a reflection, with $T_\tau$ interpreted as a 5-fold rotation. With 60 elements, we can generate the group with a minimal generating set of two elements but it is more helpful to also consider three as we have done previously, with $C_\tau=S_\tau T_\tau$. We notice that the domain shown in Figure \ref{fig:G5} appears to be missing some sections (it is no longer symmetric around the cusps). This is due to the equivalence of some points of the complex plane (brought on by  $T_\tau^N=e$). The stabilisers are compiled in Table~\ref{ta:G5}, where the stabilisers have equivalent values within the boundary of  the domain shown in Fig.~\ref{fig:G5}, given by\footnote{In Eqs.\eqref{eq:quivsA5}, most of the equivalencies are due to the redundancy between the outer-most boundaries (that is, $\Re(\tau)~=~\pm 5/2$). The only exception is $-0.4 = 0.6$ (the remaining follow trivially), which can be shown to be equivalent by making use of $\gamma_1 = \begin{pmatrix} -2 & -1 \\ 5 & 2 \end{pmatrix}$, and $\gamma_2 = \begin{pmatrix} 3 & 1 \\ 5 & 2 \end{pmatrix}$. Clearly, $\gamma_2^2=e$, and $\gamma_2 \cdot \gamma_1= T_\tau$, where $\gamma_1$, $\gamma_2 \in \Gamma_5$. Choosing $\tau = i \infty$, we have that $\gamma_1 i \infty = -2/5$, and $\gamma_2 i \infty = 3/5$. In this way, $-2/5 = \gamma_1 i \infty = \gamma_2 \cdot \gamma_2 \cdot \gamma_1 i \infty = \gamma_2 T_\tau i \infty =\gamma_2 i \infty = 3/5$. Hence, $-2/5 = 3/5$, and the remaining follow by acting $T_\tau$ on this equivalence.}:
\begin{eqnarray}
&\frac{5}{2} = -\frac{5}{2}, \qquad -\frac{12}{5}= -\frac{7}{5} = -\frac{2}{5} = \frac{3}{5} = \frac{8}{5}, \nonumber \\
&\frac{5}{2}+\frac{i}{2} = -\frac{5}{2}+\frac{i}{2}, \qquad 
\frac{5}{2}+i\frac{\sqrt{3}}{2} = -\frac{5}{2}+i\frac{\sqrt{3}}{2}, \qquad
\frac{5}{2}+\frac{i}{2 \sqrt{3}} = -\frac{5}{2}+\frac{i}{2 \sqrt{3}}, \nonumber \\
&\pm \frac{33}{14}+i\frac{\sqrt{3}}{14} = \pm\frac{23}{14}+i\frac{\sqrt{3}}{14}, \qquad 
\pm\frac{19}{14}+i\frac{\sqrt{3}}{14} = \pm\frac{9}{14}+i\frac{\sqrt{3}}{14}, \qquad
\frac{5}{14}+i\frac{\sqrt{3}}{14} = -\frac{5}{14}+i\frac{\sqrt{3}}{14}, \nonumber \\
&\frac{5}{13}+\frac{i}{13} = -\frac{5}{13}+\frac{i}{13}, \qquad
\pm\frac{18}{13}+\frac{i}{13} = \pm\frac{8}{13}+\frac{i}{13}, \qquad
\pm\frac{31}{13}+\frac{i}{13} = \pm\frac{21}{13}+\frac{i}{13}, \nonumber \\
&\frac{15}{26}+\frac{i}{26 \sqrt{3}} = \frac{15}{38}+i\frac{\sqrt{3}}{38} =-\frac{15}{38}+i\frac{\sqrt{3}}{38} = -\frac{15}{26}+\frac{i}{26 \sqrt{3}}, \nonumber \\
&\pm \frac{41}{26}+\frac{i}{26 \sqrt{3}} = \pm\frac{53}{38}+i\frac{\sqrt{3}}{38} = \pm\frac{23}{38}+i\frac{\sqrt{3}}{38}=  \pm\frac{11}{26}+\frac{i}{26 \sqrt{3}}, \nonumber \\
\label{eq:quivsA5}
&\pm\frac{91}{38}+i\frac{\sqrt{3}}{38} = \mp \frac{63}{26}+\frac{i}{26 \sqrt{3}} = \pm\frac{61}{38}+i\frac{\sqrt{3}}{38} = \pm \frac{37}{26}+\frac{i}{26 \sqrt{3}} \,.
\end{eqnarray}

In these equivalances, we stress the $\pm$ and the single $\mp$ are not interchangeable. Each equivalence featuring these symbols is a compact form encoding only two (not four) separate equivalences.

In terms of the generators $S_\tau$ and $T_\tau$, in a $T_\tau$-diagonal basis, for the triplet irreducible representations of $A_5$, we have the following representation matrices:
\begin{eqnarray}
&&  
\rho_{\mathbf{3}}(S_\tau)=\frac{1}{\sqrt{5}}
\begin{bmatrix}
 1 &~ -\sqrt{2} &~ -\sqrt{2} \\
 -\sqrt{2} &~ -\phi_g  &~ \phi_g-1 \\
 -\sqrt{2} &~ \phi_g-1 &~ -\phi_g
\end{bmatrix}\,,\quad 
\rho_{\mathbf{3}}(T_\tau)=
\begin{bmatrix}
 1 &~ 0 &~ 0 \\
 0 &~ \omega_{5}  &~ 0 \\
 0 &~ 0 &~ \omega_{5} ^4
\end{bmatrix}\,,~
\nonumber\\
&& \rho_{\mathbf{3}^{\prime}}(S_\tau)=\frac{1}{\sqrt{5}}
\begin{bmatrix}
 -1 &~ \sqrt{2} &~ \sqrt{2} \\
 \sqrt{2} &~ 1-\phi_g &~ \phi_g  \\
 \sqrt{2} &~ \phi_g  &~ 1-\phi_g
\end{bmatrix}\,, ~\quad
\rho_{\mathbf{3}^{\prime}}(T_\tau)=
\begin{bmatrix}
 1 &~ 0 &~ 0 \\
 0 &~ \omega_{5} ^2 &~ 0 \\
 0 &~ 0 &~ \omega_{5} ^3
\end{bmatrix} \,,
\end{eqnarray}
where $\phi_g = \frac{(1+\sqrt{5})}{2}$.

For the triplets of $\Gamma_5$ ($A_5$), the modular forms at stabilisers of the generators are:
\begin{eqnarray}
&&Y_{\mathbf{3}}(\tau_{S_\tau}) \propto
\begin{bmatrix} -2 \phi_g \\ \sqrt{2} \\ \sqrt{2} \end{bmatrix} \,,~
x \begin{bmatrix} \phi_g-1 \\ \sqrt{2} \\ 0 \end{bmatrix} + 
y \begin{bmatrix} \phi_g-1 \\ 0 \\ \sqrt{2} \end{bmatrix}\,, \quad
Y_{\mathbf{3}}(\tau_{T_\tau}) \propto
\begin{bmatrix} 1 \\ 0 \\ 0 \end{bmatrix} \,,~
\begin{bmatrix} 0 \\ 1 \\ 0 \end{bmatrix} \,,~
\begin{bmatrix} 0 \\ 0 \\ 1 \end{bmatrix} \,,
\nonumber\\
&&Y_{\mathbf{3}'}(\tau_{S_\tau}) \propto
\begin{bmatrix} 2 \phi_g-2 \\ \sqrt{2} \\ \sqrt{2} \end{bmatrix} \,,~
x \begin{bmatrix} -\phi_g \\ \sqrt{2} \\ 0 \end{bmatrix} + 
y \begin{bmatrix} -\phi_g \\ 0 \\ \sqrt{2} \end{bmatrix} \,,\;\,\quad
Y_{\mathbf{3'}}(\tau_{T_\tau}) \propto
\begin{bmatrix} 1 \\ 0 \\ 0 \end{bmatrix} \,,~
\begin{bmatrix} 0 \\ 1 \\ 0 \end{bmatrix} \,,~
\begin{bmatrix} 0 \\ 0 \\ 1 \end{bmatrix} \,,
\end{eqnarray}
where (again) $x,y$ are placeholder normalization factors that can be found (for a specified weight).

{
\centering
\begin{longtable}{c|c|c}\hline
	& $\gamma$ & $\tau_\gamma$ \\ \hline \hline
&&\\[-2ex]
& 
 $T_\tau^3 C_\tau T_\tau C_\tau$ & $-\frac{3}{2}+\frac{i}{2 \sqrt{3}},\frac{19}{14}+\frac{i \sqrt{3}}{14}$ \\
&
 $C_\tau T_\tau C_\tau T_\tau^2$ & $-\frac{3}{2}+\frac{i}{2 \sqrt{3}}, \frac{19}{14}+\frac{i \sqrt{3}}{14}$ \\
&
 $C_\tau T_\tau C_\tau T_\tau C_\tau$ & $-\frac{1}{2}+\frac{i}{2 \sqrt{3}},\frac{33}{14}+\frac{i \sqrt{3}}{14}$ \\
&
 $C_\tau T_\tau^2 C_\tau$ & $-\frac{1}{2}+\frac{i}{2 \sqrt{3}}, \frac{33}{14}+\frac{i \sqrt{3}}{14}$ \\
&
 $T_\tau^3 C_\tau T_\tau^2$ & $\frac{15}{26}+\frac{i}{26 \sqrt{3}}, \frac{5}{2}+\frac{i \sqrt{3}}{2}$ \\
&
 $T_\tau^2 C_\tau T_\tau$ & $\frac{15}{26}+\frac{i}{26 \sqrt{3}}, \frac{5}{2}+\frac{i \sqrt{3}}{2}$ \\
&
 $T_\tau S_\tau$ & $-\frac{37}{26}+\frac{i}{26  \sqrt{3}}, \frac{1}{2}+\frac{i \sqrt{3}}{2}$ \\
&
 $T_\tau C_\tau S_\tau$ & $-\frac{37}{26}+\frac{i}{26  \sqrt{3}}, \frac{1}{2}+\frac{i \sqrt{3}}{2}$ \\
&
 $C_\tau T_\tau^2 C_\tau S_\tau$ & $-\frac{23}{14}+\frac{i \sqrt{3}}{14}, \frac{1}{2}+\frac{i}{2 \sqrt{3}}$ \\
\multirow{2}{*}{$\mathcal{C}_2$} 
&
 $T_\tau^2 C_\tau T_\tau C_\tau$ & $-\frac{23}{14}+\frac{i \sqrt{3}}{14}, \frac{1}{2}+\frac{i}{2 \sqrt{3}}$ \\
&
 $C_\tau^2 T_\tau^2$ & $-\frac{3}{2}+\frac{i \sqrt{3}}{2},\frac{41}{26}+\frac{i}{26 \sqrt{3}}$ \\
&
 $T_\tau^3 C_\tau$ & $-\frac{3}{2}+\frac{i \sqrt{3}}{2},\frac{41}{26}+\frac{i}{26 \sqrt{3}}$ \\
&
 $C_\tau T_\tau^2 C_\tau T_\tau$ & $-\frac{9}{14}+\frac{i \sqrt{3}}{14}, \frac{3}{2}+\frac{i}{2 \sqrt{3}}$ \\
&
 $T_\tau^3 C_\tau T_\tau S_\tau$ & $-\frac{9}{14}+\frac{i \sqrt{3}}{14},\frac{3}{2}+\frac{i}{2 \sqrt{3}}$ \\
&
 $T_\tau^3 C_\tau S_\tau$ & $-\frac{41}{26}+\frac{i}{26 \sqrt{3}}, \frac{3}{2}+\frac{i \sqrt{3}}{2}$ \\
&
 $T_\tau C_\tau T_\tau^2$ & $-\frac{41}{26}+\frac{i}{26 \sqrt{3}}, \frac{3}{2}+\frac{i \sqrt{3}}{2}$ \\
&
 $C_\tau$ & $-\frac{1}{2}+\frac{i \sqrt{3}}{2},\frac{91}{38}+\frac{i \sqrt{3}}{38}$ \\
&
 $C_\tau^2$ & $-\frac{1}{2}+\frac{i \sqrt{3}}{2},\frac{91}{38}+\frac{i \sqrt{3}}{38}$ \\
&
 $T_\tau C_\tau T_\tau^2 C_\tau$ & $\frac{5}{14}+\frac{i \sqrt{3}}{14},\frac{5}{2}+\frac{i}{2 \sqrt{3}}$ \\
&
 $C_\tau T_\tau C_\tau T_\tau S_\tau$ & $\frac{5}{14}+\frac{i \sqrt{3}}{14},\frac{5}{2}+\frac{i}{2 \sqrt{3}}$ \\[-2ex]
 &&\\ \hline \hline
&&\\[-2ex]
& 
$C_\tau^2 T_\tau$ & $-1+i,\frac{46}{29}+\frac{i}{29}$ \\
&
$T_\tau^2 C_\tau T_\tau^2 C_\tau S_\tau$ & $-\frac{8}{13}+\frac{i}{13},\frac{3}{2}+\frac{i}{2}$ \\
&
$C_\tau T_\tau^2 C_\tau T_\tau S_\tau$ & $-\frac{12}{5}+\frac{i}{5},\frac{2}{5}+\frac{i}{5}$ \\
&
$T_\tau^2 C_\tau T_\tau^2$ & $-\frac{12}{29}+\frac{i}{29},2+i$ \\
&
$S_\tau$ & $-\frac{70}{29}+\frac{i}{29},i$ \\
&
$C_\tau T_\tau^2 C_\tau T_\tau C_\tau$ & $-\frac{2}{5}+\frac{i}{5},\frac{12}{5}+\frac{i}{5}$ \\
&
$C_\tau T_\tau C_\tau T_\tau^2 C_\tau$ & $-\frac{3}{5}+\frac{i}{5}, \frac{8}{5}+\frac{i}{5}$ \\
$\mathcal{C}_3$
&
$T_\tau^2 C_\tau T_\tau^2 C_\tau T_\tau$ & $\frac{5}{13}+\frac{i}{13},\frac{5}{2}+\frac{i}{2}$ \\
&
$C_\tau^2 T_\tau^2 C_\tau T_\tau$ & $-\frac{3}{2}+\frac{i}{2},\frac{18}{13}+\frac{i}{13}$ \\
&
$T_\tau C_\tau T_\tau S_\tau$ & $-\frac{21}{13}+\frac{i}{13}, \frac{1}{2}+\frac{i}{2}$ \\
&
$C_\tau T_\tau C_\tau T_\tau^2 C_\tau T_\tau$ & $-\frac{7}{5}+\frac{i}{5},\frac{7}{5}+\frac{i}{5}$ \\
&
$C_\tau T_\tau C_\tau$ & $-\frac{1}{2}+\frac{i}{2},\frac{31}{13}+\frac{i}{13}$ \\
&
$T_\tau^3 C_\tau T_\tau$ & $-2+i,\frac{17}{29}+\frac{i}{29}$ \\
&
$T_\tau^2 C_\tau S_\tau$ & $-\frac{41}{29}+\frac{i}{29}, 1+i$ \\
&
$T_\tau C_\tau T_\tau^2 C_\tau T_\tau S_\tau$ & $-\frac{8}{5}+\frac{i}{5}, \frac{3}{5}+\frac{i}{5}$ \\[-2ex] 
&&\\\hline \hline
&&\\[-2ex]
&
$T_\tau$ & $i \infty, \frac{8}{5}$ \\
&
$T_\tau^4$ & $i \infty, \frac{8}{5}$ \\
&
$T_\tau C_\tau$ & $0,\frac{5}{2}$ \\
&
$C_\tau S_\tau$ & $0,\frac{5}{2}$ \\
&
$C_\tau^2 T_\tau S_\tau$ & $-\frac{1}{2},2$ \\
\multirow{2}{*}{$\mathcal{C}_4$} 
&
$T_\tau C_\tau T_\tau$ & $-\frac{1}{2},2$ \\
&
$T_\tau^3 S_\tau$ & $-1,\frac{3}{2}$ \\
&
$C_\tau T_\tau$ & $-1,\frac{3}{2}$ \\
&
$T_\tau^2 S_\tau$ & $-\frac{3}{2},1$ \\
&
$C_\tau T_\tau^2$ & $-\frac{3}{2},1$ \\
&
$C_\tau T_\tau C_\tau S_\tau$ & $-2,\frac{1}{2}$ \\
&
$T_\tau^2 C_\tau$ & $-2,\frac{1}{2}$ \\[-2ex] 
&&\\ \hline \hline
&&\\[-2ex]
&
$T_\tau^2$ & $i \infty, \frac{8}{5}$ \\
&
$T_\tau^3$ & $i \infty, \frac{8}{5}$ \\
&
$C_\tau T_\tau S_\tau$ & $0,\frac{5}{2}$ \\
&
$T_\tau C_\tau T_\tau C_\tau$ & $0,\frac{5}{2}$ \\
&
$C_\tau^2 T_\tau^2 C_\tau T_\tau S_\tau$ & $-\frac{1}{2},2$ \\
\multirow{2}{*}{$\mathcal{C}_5$} 
&
$T_\tau C_\tau T_\tau^2 C_\tau T_\tau$ & $-\frac{1}{2},2$\\ 
&
$C_\tau^2 T_\tau^2 C_\tau$ & $-1,\frac{3}{2}$ \\
&
$C_\tau T_\tau C_\tau T_\tau$ & $-1,\frac{3}{2}$ \\
&
$T_\tau^2 C_\tau T_\tau S_\tau$ & $-\frac{3}{2},1$ \\
&
$T_\tau C_\tau T_\tau^2 C_\tau S_\tau$ & $-\frac{3}{2},1$ \\
&
$T_\tau^2 C_\tau T_\tau^2 C_\tau$ & $-2,\frac{1}{2}$ \\
&
$C_\tau T_\tau C_\tau T_\tau^2 C_\tau S_\tau$ & $-2,\frac{1}{2}$ \\[-2ex] 
&&\\\hline
\caption{\label{ta:G5} The non-identity elements of $\Gamma_5$ and respective stabilisers.}
\end{longtable}
}

\section{Conclusion \label{sec:conc}}

In this work, we have described and employed an algorithm for identifying stabilisers $\tau_\gamma$ for finite modular groups. We used the algorithm to find all inequivalent stabilisers for each group element $\gamma \in \Gamma_{2,3,4,5}$, i.e. for finite modular groups with $N$ up to 5. We have shown the stabilisers in the domains of the respective modular symmetries, in the upper complex plane, and the tables \ref{ta:G2}-\ref{ta:G5} list our findings. The stabilisers listed are complete in the sense that we present all inequivalent stabilisers. Nevertheless, we note that these have infinite multiplicities in the upper complex plane, but we show the explicit multiplicities in the domains shown within the figures. Given that each group element by itself generates a specific cyclic subgroup of the finite modular symmetry, our work provides stabilisers for each of these cyclic subgroups, and is therefore useful to applications of finite modular symmetries that are broken to residual subgroups. In particular, this work is intended to assist model-building efforts when finite modular symmetries are used as flavour symmetries, to account for fermion masses and mixing.

\section*{Acknowledgements}
IdMV acknowledges
funding from Funda\c{c}\~{a}o para a Ci\^{e}ncia e a Tecnologia (FCT) through the
contract IF/00816/2015 and was supported in part by the National Science Center, Poland, through the HARMONIA project under contract UMO-2015/18/M/ST2/00518 (2016-2019), and by FCT through projects CFTP-FCT Unit 777 (UID/FIS/00777/2019), CERN/FIS-PAR/0004/2017 and PTDC/FIS-PAR/29436/2017 which are partially funded through POCTI (FEDER), COMPETE, QREN and EU.
 The work of ML is funded by
Funda\c{c}\~{a}o para a Ci\^{e}ncia e Tecnologia-FCT Grant
No.PD/BD/150488/2019, in the framework of the Doctoral Programme
IDPASC-PT.  
YLZ acknowledges the STFC Consolidated Grant ST/L000296/1 and the European Union's Horizon 2020 Research and Innovation programme under Marie Sk\l{}odowska-Curie grant agreements Elusives ITN No.\ 674896 and InvisiblesPlus RISE No.\ 690575.


\begin{thebibliography}{99}

\bibitem{Altarelli:2005yp}
  G.~Altarelli and F.~Feruglio,
  Nucl.\ Phys.\ B {\bf 720} (2005) 64
  [hep-ph/0504165].


\bibitem{Altarelli:2005yx}
  G.~Altarelli and F.~Feruglio,
  Nucl.\ Phys.\ B {\bf 741} (2006) 215
  [hep-ph/0512103].


\bibitem{King:2017guk}
  S.~F.~King,
  Prog.\ Part.\ Nucl.\ Phys.\  {\bf 94} (2017) 217
  [arXiv:1701.04413 [hep-ph]].


\bibitem{Xing:2019vks}
  Z.~z.~Xing,
  Phys.\ Rept.\  {\bf 854} (2020) 1
  [arXiv:1909.09610 [hep-ph]].


\bibitem{Feruglio:2019ktm}
  F.~Feruglio and A.~Romanino,
  arXiv:1912.06028 [hep-ph].


\bibitem{Ferrara:1989bc}
  S.~Ferrara, D.~Lust, A.~D.~Shapere and S.~Theisen,
  Phys.\ Lett.\ B {\bf 225} (1989) 363.


\bibitem{Ferrara:1989qb}
  S.~Ferrara, .D.~Lust and S.~Theisen,
  Phys.\ Lett.\ B {\bf 233} (1989) 147.


\bibitem{Feruglio:2017spp}
  F.~Feruglio,
  arXiv:1706.08749 [hep-ph].


\bibitem{Criado:2018thu}
  J.~C.~Criado and F.~Feruglio,
  SciPost Phys.\  {\bf 5} (2018) no.5,  042
  [arXiv:1807.01125 [hep-ph]].


\bibitem{deMedeirosVarzielas:2019cyj}
  I.~de Medeiros Varzielas, S.~F.~King and Y.~L.~Zhou,
  Phys.\ Rev.\ D {\bf 101} (2020) no.5,  055033
  [arXiv:1906.02208 [hep-ph]].


\bibitem{Kobayashi:2018vbk}
  T.~Kobayashi, K.~Tanaka and T.~H.~Tatsuishi,
  Phys.\ Rev.\ D {\bf 98} (2018) no.1,  016004
  [arXiv:1803.10391 [hep-ph]].


\bibitem{Kobayashi:2018wkl}
  T.~Kobayashi, Y.~Shimizu, K.~Takagi, M.~Tanimoto, T.~H.~Tatsuishi and H.~Uchida,
  Phys.\ Lett.\ B {\bf 794} (2019) 114
  [arXiv:1812.11072 [hep-ph]].


\bibitem{Kobayashi:2018scp}
  T.~Kobayashi, N.~Omoto, Y.~Shimizu, K.~Takagi, M.~Tanimoto and T.~H.~Tatsuishi,
  JHEP {\bf 1811} (2018) 196
  [arXiv:1808.03012 [hep-ph]].


\bibitem{Okada:2018yrn}
  H.~Okada and M.~Tanimoto,
  Phys.\ Lett.\ B {\bf 791} (2019) 54
  [arXiv:1812.09677 [hep-ph]].


\bibitem{Novichkov:2018yse}
  P.~P.~Novichkov, S.~T.~Petcov and M.~Tanimoto,
  Phys.\ Lett.\ B {\bf 793} (2019) 247
  [arXiv:1812.11289 [hep-ph]].


\bibitem{Ding:2019zxk}
  G.~J.~Ding, S.~F.~King and X.~G.~Liu,
  JHEP {\bf 1909} (2019) 074
  [arXiv:1907.11714 [hep-ph]].


\bibitem{Zhang:2019ngf}
  D.~Zhang,
  Nucl.\ Phys.\ B {\bf 952} (2020) 114935
  [arXiv:1910.07869 [hep-ph]].


\bibitem{Wang:2019xbo}
  X.~Wang,
  Nucl.\ Phys.\ B {\bf 957} (2020) 115105
  [arXiv:1912.13284 [hep-ph]].


\bibitem{Penedo:2018nmg}
  J.~T.~Penedo and S.~T.~Petcov,
  Nucl.\ Phys.\ B {\bf 939} (2019) 292
  [arXiv:1806.11040 [hep-ph]].


\bibitem{Novichkov:2018ovf}
  P.~P.~Novichkov, J.~T.~Penedo, S.~T.~Petcov and A.~V.~Titov,
  JHEP {\bf 1904} (2019) 005
  [arXiv:1811.04933 [hep-ph]].


\bibitem{King:2019vhv}
  S.~F.~King and Y.~L.~Zhou,
  Phys.\ Rev.\ D {\bf 101} (2020) no.1,  015001
  [arXiv:1908.02770 [hep-ph]].


\bibitem{Wang:2019ovr}
  X.~Wang and S.~Zhou,
  JHEP {\bf 2005} (2020) 017
  [arXiv:1910.09473 [hep-ph]].


\bibitem{Wang:2020dbp}
  X.~Wang,
  arXiv:2007.05913 [hep-ph].


\bibitem{Novichkov:2018nkm}
  P.~P.~Novichkov, J.~T.~Penedo, S.~T.~Petcov and A.~V.~Titov,
  JHEP {\bf 1904} (2019) 174
  [arXiv:1812.02158 [hep-ph]].


\bibitem{Ding:2019xna}
  G.~J.~Ding, S.~F.~King and X.~G.~Liu,
  Phys.\ Rev.\ D {\bf 100} (2019) no.11,  115005
  [arXiv:1903.12588 [hep-ph]].


\bibitem{Ding:2020msi}
G.~J.~Ding, S.~F.~King, C.~C.~Li and Y.~L.~Zhou,
JHEP \textbf{08} (2020), 164
[arXiv:2004.12662 [hep-ph]].


\bibitem{Liu:2019khw}
  X.~G.~Liu and G.~J.~Ding,
  JHEP {\bf 1908} (2019) 134
  [arXiv:1907.01488 [hep-ph]].


\bibitem{Liu:2020akv}
  X.~G.~Liu, C.~Y.~Yao and G.~J.~Ding,
  arXiv:2006.10722 [hep-ph].


\bibitem{Novichkov:2020eep}
  P.~P.~Novichkov, J.~T.~Penedo and S.~T.~Petcov,
  arXiv:2006.03058 [hep-ph].


\bibitem{Liu:2020msy}
  X.~G.~Liu, C.~Y.~Yao, B.~Y.~Qu and G.~J.~Ding,
  arXiv:2007.13706 [hep-ph].


\bibitem{Okada:2019uoy}
  H.~Okada and M.~Tanimoto,
  arXiv:1905.13421 [hep-ph].


\bibitem{King:2020qaj}
S.~J.~D.~King and S.~F.~King,
JHEP \textbf{09} (2020), 043
[arXiv:2002.00969 [hep-ph]].

\bibitem{deAnda:2018ecu}
  F.~J.~de Anda, S.~F.~King and E.~Perdomo,
  Phys.\ Rev.\ D {\bf 101} (2020) no.1,  015028
  [arXiv:1812.05620 [hep-ph]].


\bibitem{Gui-JunDing:2019wap}
  G.~J.~Ding, S.~F.~King, X.~G.~Liu and J.~N.~Lu,
  JHEP {\bf 1912} (2019) 030
  [arXiv:1910.03460 [hep-ph]].


\bibitem{deAdelhartToorop:2011re}
  R.~de Adelhart Toorop, F.~Feruglio and C.~Hagedorn,
  Nucl.\ Phys.\ B {\bf 858} (2012) 437
  [arXiv:1112.1340 [hep-ph]].
  
\end{thebibliography}
\end{document}